\documentclass[a4paper,11pt]{article}
\usepackage{jinstpub}
\usepackage{lineno}
\usepackage{graphicx}
\usepackage{subcaption} 
\usepackage{comment}
\usepackage{subcaption}
\graphicspath{{figures/}}

\title{Liquid argon purification and purity monitoring: apparatus and first results}

\author[a,1]{W.-Z.~Wei,\note{Corresponding author.}}
\author[a,2]{I.W.~Jaidee,\note{Current address: Department of Physics, Harvard University, Cambridge, MA 02138, USA}}
\author[a,3]{S.~Dockal,\note{Current address: Department of Physics, Syracuse University, Syracuse, NY 13244, USA}}
\author[a]{V.~Tsvetkova,}
\author[a]{G.~Bui,}
\author[a]{T.~Chen~Lin,}
\author[a]{L.~Epstein,}
\author[a]{A.~Faubus,}
\author[a]{N.M.T.~Hambraeus,}
\author[a]{S.B.~Lyon,}
\author[a]{D.~Lopez,}
\author[a]{N.~McGee}
\author[a]{P.J.~Migden,}
\author[a]{C.~Nicollin,}
\author[a]{M.~Unnithan,}
\author[b]{J.~Asaadi,}
\author[a]{J.B.R.~Battat}

\affiliation[a]{Department of Physics and Astronomy, Wellesley College, \\ Wellesley, MA 02481, USA}
\affiliation[b]{Department of Physics, University of Texas at Arlington, \\ Arlington, TX 76019, USA}

\emailAdd{ww102@wellesley.edu}

\abstract{We report results from a 13-liter purified liquid argon test stand at Wellesley College in Massachusetts, USA. The system includes a single-pass liquid-phase purification column that can achieve sub-part-per-billion impurity levels. The liquid argon purity is measured with a double-gridded purity monitor to assess the electron lifetime. Measurements demonstrate an O$_2$-equivalent impurity concentration of $0.21 \pm 0.05$\,ppb (stable over a 20-hour duration), corresponding to an electron lifetime of 1.6\,ms at a drift field of 500\,V/cm. This test stand supports ongoing detector R\&D on charge and light readout for liquid argon time projection chambers, such as Q-Pix and other cold electronics systems.}

\keywords{LArTPC, Cryogenic detectors, Noble liquid detectors, Instrumentation and methods for detectors}

\arxivnumber{2604.21307} % Only if you have one

\begin{document}
\maketitle
\flushbottom
\section{Introduction}

Liquid argon time projection chambers (LArTPCs) have become a prominent technology for neutrino and rare-event physics due to their excellent spatial resolution, calorimetric capabilities, and scalability~\cite{rubbia1977, Majumdar:2021llu, McDonald:2024osu}. They are used in several current and future large-scale experiments, including dark matter searches such as DarkSide-20k~\cite{DarkSide-20k:2017zyg}, as well as neutrino experiments such as ICARUS~\cite{Pia:2025ydd}, MicroBooNE~\cite{microboone2017}, SBND (Short-Baseline Near Detector)~\cite{SBND:2025lha}, and the Deep Underground Neutrino Experiment (DUNE)~\cite{dune2020a}.

In a LArTPC, charged particles traversing liquid argon (LAr) produce ionization electrons and scintillation light. Under an applied electric field, the ionization electrons drift toward a readout plane, where their 2D arrival positions combined with the differences in arrival times provide a full 3D reconstruction of the ionization distribution within the LArTPC. When the absolute drift time is known from the prompt scintillation light, the absolute position of the interaction vertex along the drift direction can also be measured, allowing full-volume fiducialization.  In addition to imaging the event topology, the collected ionization charge provides a measurement of the energy deposited by the particle along its path. After correcting for effects such as electron attenuation and recombination, the charge per unit length can be converted to deposited energy, enabling calorimetric energy reconstruction and particle identification. 

The successful operation of LArTPCs critically depends on maintaining ultra-high purity of the LAr to prevent degradation of scintillation and ionization yields~\cite{Vogl:2023hbg}. Nitrogen and oxygen quench argon excimer states and thereby reduce the scintillation light yield~\cite{WArP:2008rgv, WArP:2008dyo}, while oxygen and water absorb vacuum ultraviolet (VUV) photons~\cite{WArP:2008dyo, Onaka1968}. In addition, drifting electrons may attach to electronegative impurities (primarily H$_2$O and O$_2$), attenuating the ionization signal~\cite{Bakale1976}. The electron lifetime is commonly used as a performance benchmark. At a typical drift field of 500\,V/cm\footnote{A drift field of 500\,V/cm is commonly chosen in LArTPCs to balance scintillation and ionization signal yields.}, the electron drift velocity in LAr is 1.6\,mm/$\mu$s~\cite{lar_properties}. For large-scale detectors with drift distances of several meters, this corresponds to electron transit times of many milliseconds. Achieving adequate electron drift lifetimes requires impurity concentrations below a part-per-billion (ppb), which is several orders of magnitude lower than commercially available in ultra-high-purity (UHP) LAr. Consequently, ensuring and continuously monitoring argon purity remains a central priority in LArTPC operation.

 Several LAr purification and purity monitoring systems have been built, including the Liquid Argon Purity Demonstrator (LAPD)~\cite{Adamowski:2014daa} and Luke (also known as the Materials Test Stand (MTS))~\cite{Andrews:2009zza} at Fermilab, ARGONTUBE~\cite{Badhrees:2012zz, Ereditato:2013xaa} at the University of Bern, and more recent large-scale R\&D prototypes such as CAPTAIN~\cite{CAPTAIN:2020pup} at Los Alamos and protoDUNE-SP~\cite{DUNE:2025wjg} at CERN. These facilities generally employ continuous LAr recirculation through dedicated filters to remove impurities, which allows efficient purification of large volumes, along with purity monitors to track the electron lifetime. In the most widely used purity monitor design, developed by the ICARUS collaboration~\cite{CARUGNO1990580}, ultraviolet (UV) light liberates electrons from a photocathode via the photoelectric effect. 
A uniform electric field drives the resulting electrons toward an anode, and the electron lifetime is determined from the attenuation of the electron signal. The design and operation of the UV-based purity monitor (UV-PrM) is described in detail in Ref.~\cite{fogarty_prm_2023}.
In this work, we report on the design, construction, operation and performance of a LAr test stand at Wellesley College that uses a single-pass purifier to remove the electronegative contaminants O$_2$ and H$_2$O, and a UV-PrM to measure and monitor the LAr purity. 
This test stand supports ongoing detector R\&D for novel LArTPC charge and light readout technologies such as Q-Pix~\cite{Nygren:2018QPix, Asaadi:2024Pix}. 
 
 This paper is organized as follows. Section~\ref{sec:system} details the design of the purifier and UV-PrM. Section~\ref{sec:procedure} discusses our experimental procedures. Section~\ref{sec:sys_perf} describes the data acquisition system and presents our experimental results. Finally, section~\ref{sec:summary} summarizes the results, lessons learned, and future plans.

\section{System overview}
\label{sec:system}

\begin{figure}[t]
    \centering
    \includegraphics[width=\textwidth]{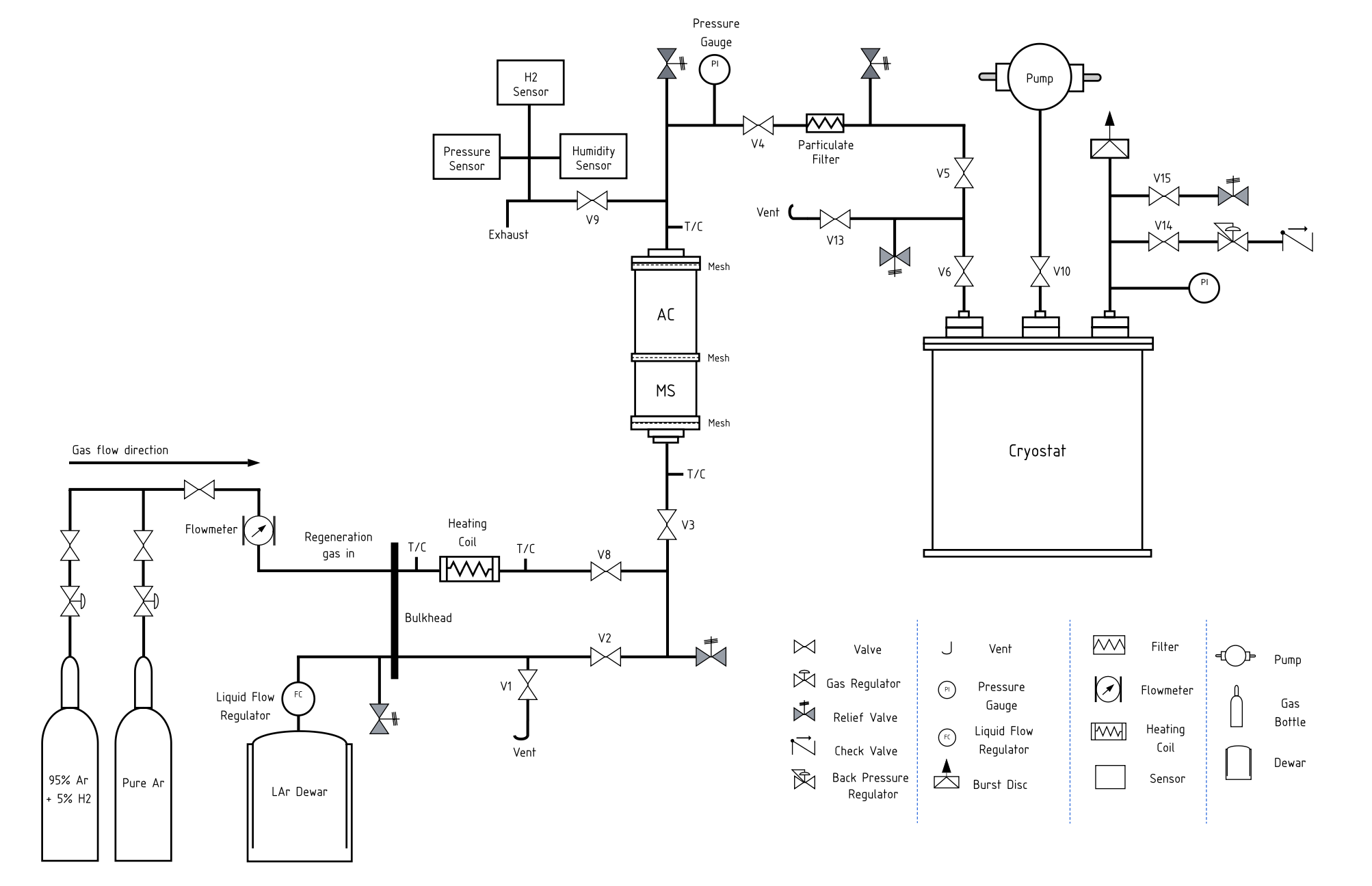}
    \vspace{0.5em}
    \includegraphics[width=\textwidth]{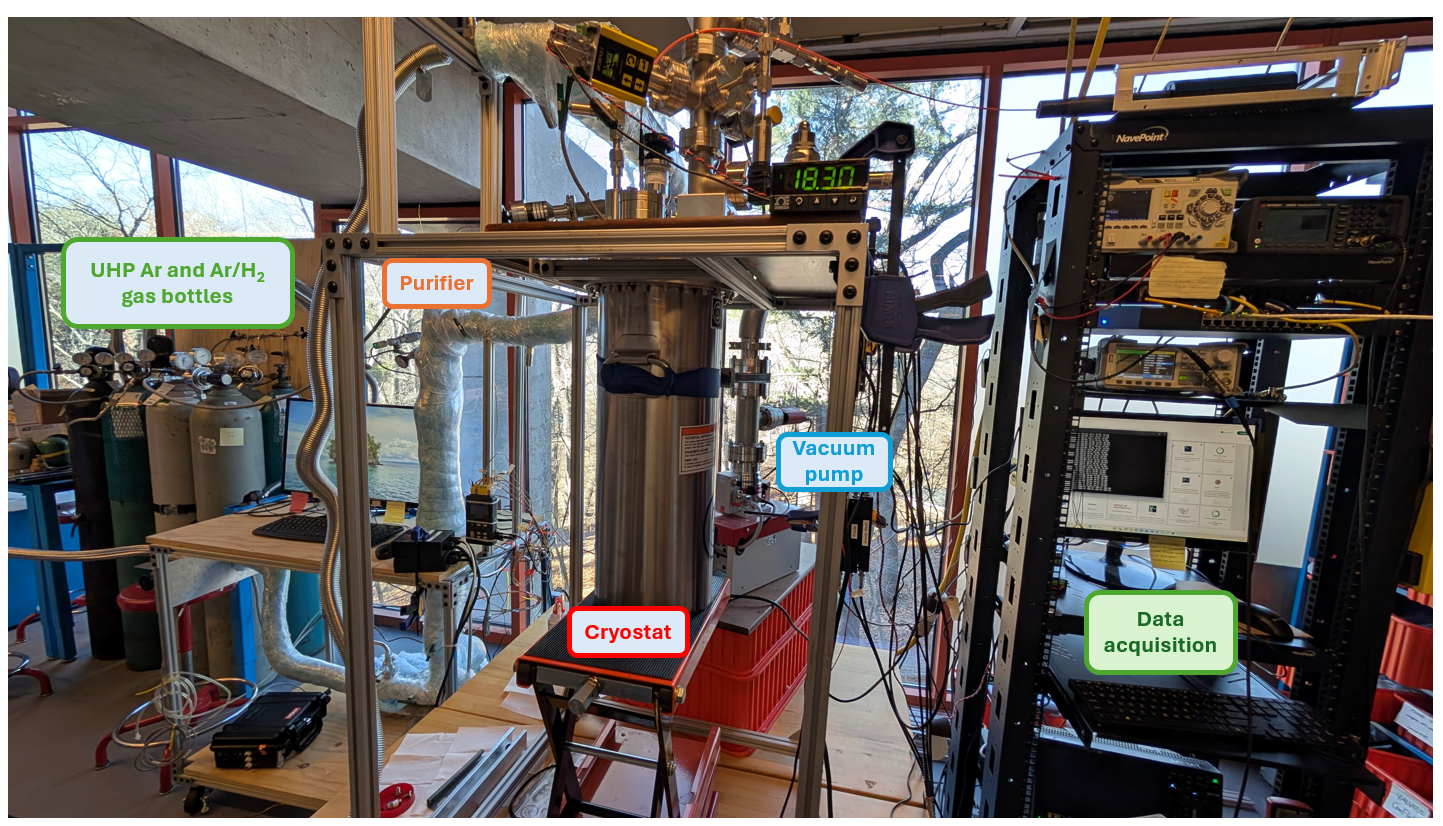}
    \caption{\textit{Top:} P\&ID diagram of the system. \textit{Bottom:} Photograph of the experimental setup, showing the gas bottles used for purifier regeneration, the purification column, the vacuum pump station, the cryostat housing the purity monitor, and the data acquisition system. The xenon flash lamp and control electronics are located at the top of the electronics rack. LAr flows upward through the purifier column before entering the cryostat.}
    \label{fig:setup}
\end{figure}

Figure~\ref{fig:setup} shows the piping and instrumentation diagram (P\&ID) and photograph of the system, which consists of a single-pass purifier, a cryostat housing a UV-PrM along with associated hardware and a vacuum pumping station. A slow-control and data acquisition system allows for remote monitoring of the apparatus.

During purification, LAr flows from a dewar of commercially available ultra high purity liquid argon (typically with oxygen and water contamination levels of 0.1 and 1\,ppm, respectively) through the purifier, and into the cryostat that houses the UV-PrM. Each sub-system is described in more detail below.

\subsection{The liquid-phase single-pass purifier}
\label{subsec:purifier}
\begin{figure}[t]
    \centering
    % First subfigure
    \begin{subfigure}[b]{0.45\textwidth}
        \centering
        \includegraphics[width=\textwidth]{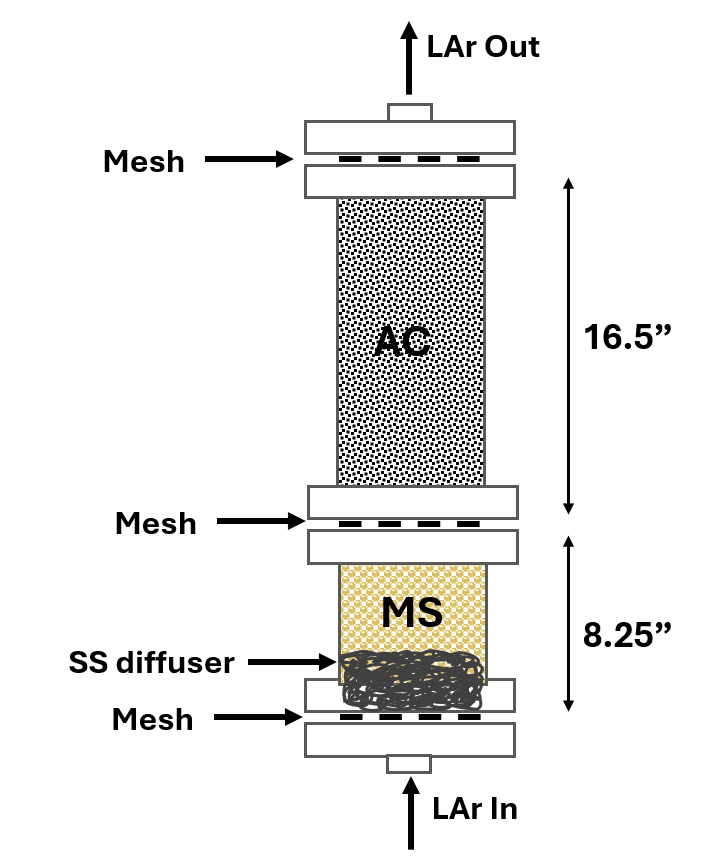}
    \end{subfigure}
    \hfill
    % Second subfigure
    \begin{subfigure}[b]{0.5\textwidth}
        \centering
        \includegraphics[width=\textwidth]{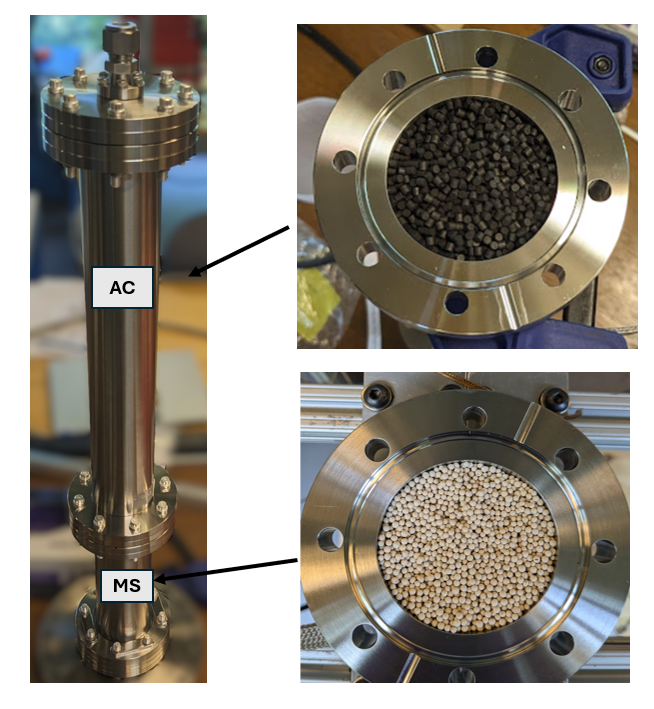}
    \end{subfigure}
    \caption{\textit{Left:} Schematic diagram of the purification column. \textit{Right:} Photograph of the purifier with the interior view of the AC catalyst pellets, which appear black due to oxidation (copper-colored after regeneration), and the MS beads.}
    \label{fig:purifier}
\end{figure}

The single-pass liquid-phase purifier consists of two filters in series, as shown in Fig.~\ref{fig:purifier}. The first filter contains 0.3\,kg of 4\,\AA{} molecular sieve (MS) in a 8.25\,inch-long 4.5" ConFlat (CF) nipple. The MS is primarily responsible for removing water, but it can also remove small amounts of nitrogen and residual oxygen. The second filter consists of 1.7\,kg of activated copper (AC) catalyst in a 16.5~inch-long 4.5" CF nipple and is used primarily for oxygen removal and, to a lesser extent, for water removal. A stainless steel (SS) mesh between the two nipples prevents filter media mixing. Throughout this paper we refer to the two filter media as the MS column and the AC column.

\begin{figure}[t]
    \centering
    % First subfigure
    \begin{subfigure}[b]{0.45\textwidth}
        \centering
        \includegraphics[width=\textwidth]{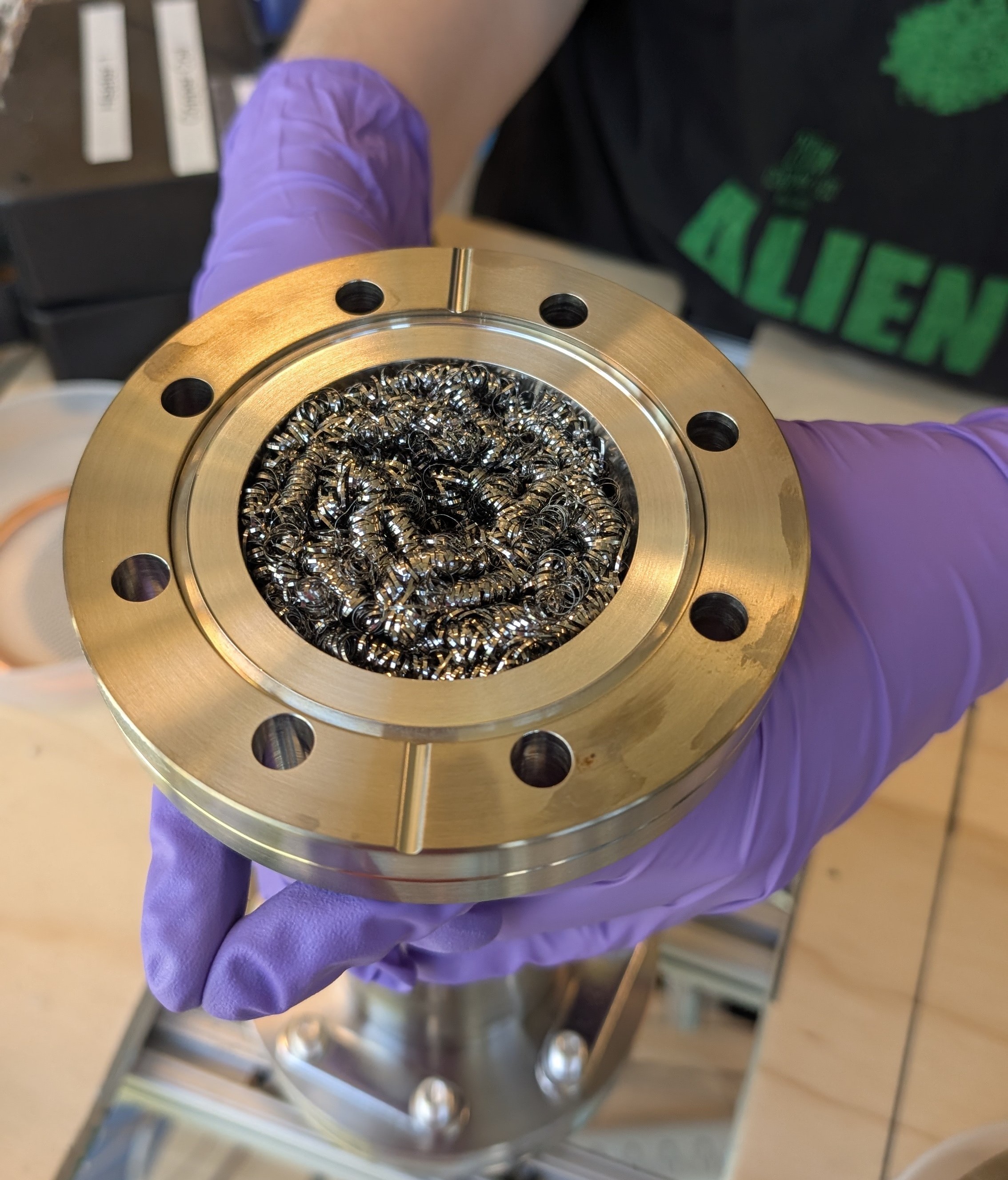}
    \end{subfigure}
    \hfill
    % Second subfigure
    \begin{subfigure}[b]{0.45\textwidth}
        \centering
        \includegraphics[width=\textwidth]{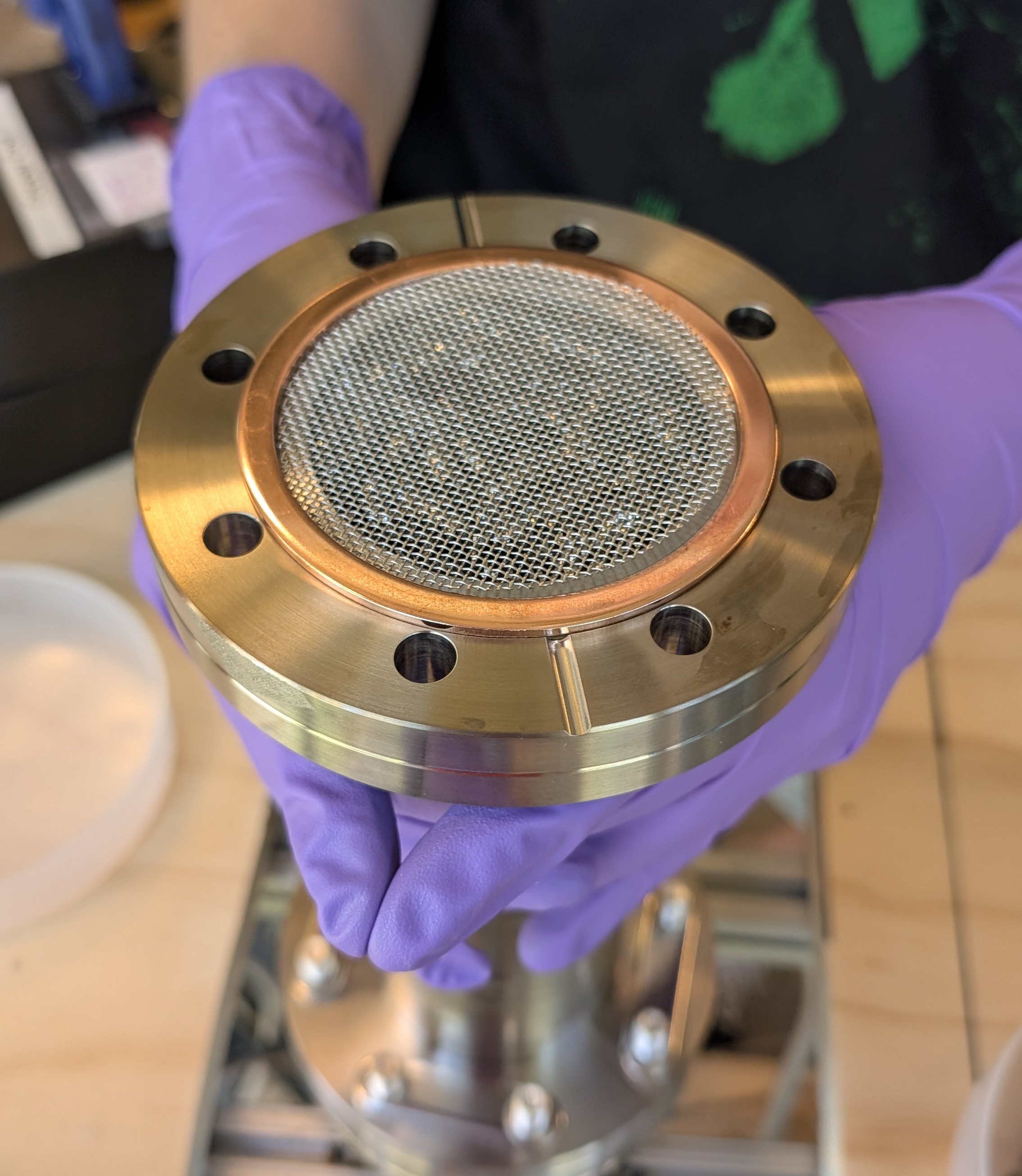}
    \end{subfigure}
    \caption{Stainless steel diffuser (left) and mesh (right) were installed at the purifier entrance, where LAr flows into the MS column.}
    \label{fig:scrubber_mesh}
\end{figure}

The purifier performance depends on several factors, including liquid-filter contact, flow rate, purifier material capacity and regeneration, as well as system cleanliness, outgassing, and thermal stability. To maximize contact between the liquid and the filter media and to prevent channeling, a SS diffuser was installed at the purifier inlet (see Fig.~\ref{fig:scrubber_mesh}, left). The diffuser, together with a controlled flow rate, helps distribute the LAr more evenly throughout the filter media, improving contact with the purifier materials. To mechanically secure the diffuser and prevent the MS beads from escaping the column, a SS mesh was placed above the diffuser (Fig.~\ref{fig:scrubber_mesh}, right).

\begin{figure}[t]
    \centering
    \includegraphics[width=\textwidth]{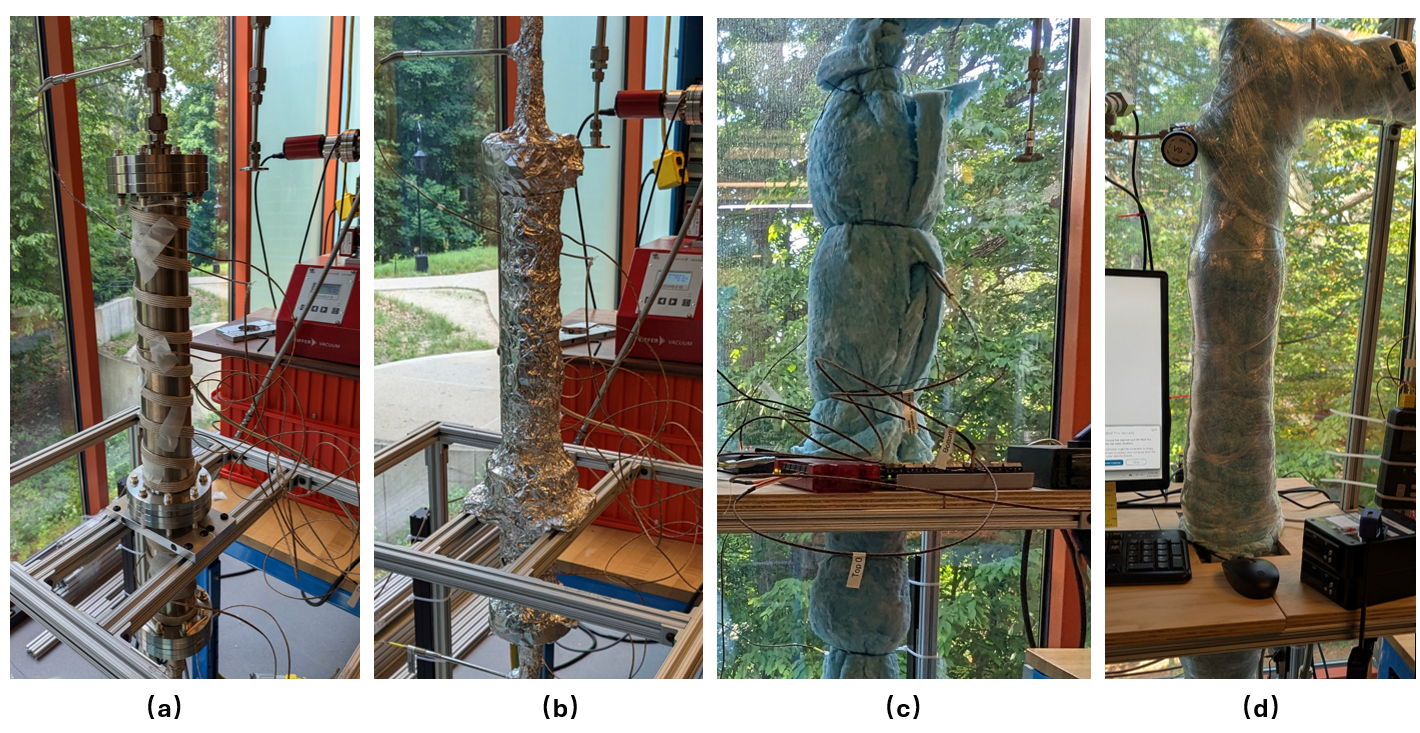}
    \caption{Purification column exterior with (a) heater tapes and thermocouples covered with (b) aluminum foil, (c) Cryo-Lite fiberglass insulation, and (d) plastic wrap as a vapor barrier.}
    \label{fig:purifier_outside}
\end{figure}

Temperature monitoring and control are critical during both regeneration and LAr purification. The purifier is instrumented with eight K-type thermocouples (see Fig.~\ref{fig:purifier_outside} (a)) to monitor the temperature at various locations during both regeneration and LAr flow: four are embedded within the column volume, and four are attached to the exterior surfaces of the two CF nipples. Also shown in panel (a) of Fig.~\ref{fig:purifier_outside} are two heater tapes wrapped around the columns, each controlled by a "bang-bang" (on–off) heater controller, which heats the column during filter regeneration (Section~\ref{subsec:regeneration}). Figure~\ref{fig:purifier_outside} also shows the two layers of insulation used to reduce heat exchange with the environment and improve thermal stability: aluminum foil (b) and Cryo-Lite fiberglass (c). The insulation serves dual purposes: during LAr purification, it minimizes heat transfer to limit boil-off and maintain uniform flow; during purifier regeneration, it allows the column to reach and sustain target temperatures efficiently, enhancing the effectiveness of the regeneration. Panel (d) of Fig.~\ref{fig:purifier_outside} shows plastic wrap applied outside the Cryo-Lite fiberglass, which serves as a moisture barrier to prevent condensation on the fiberglass insulation.

In addition to the thermocouples, heater tapes, and insulation layers, the purifier is instrumented with a pressure sensor, a humidity sensor, and a hydrogen concentration monitor (Nova Analytical Systems Model 5000-335-JK) to track the status of the regeneration process. These sensors provide information on the pressure, water content, and hydrogen content of the gas flowing through the filter media. Further details are provided in Section~\ref{subsec:regeneration}.

\subsection{The cryostat and purity monitor}
\label{sec:cryo_PrM}
The cryostat is a 13-liter cylindrical vacuum jacketed dewar (Cryofab CF7518) with an internal depth of 18\,in, an inner diameter of 7.45\,in, and a 10-inch CF flange. The cryostat is sealed with a six-port, 10-inch CF top plate (Kurt J.\ Lesker CF1000S6). The six ports on the top plate accommodate LAr and gas inlets, the vacuum pump, optical fiber for UV light injection, high-voltage and signal lines, resistive  temperature detectors (RTD) to monitor the LAr level, and multiple pressure gauges and pressure management devices including a back-pressure regulator with a check valve, a pressure relief valve, a burst disk, and several pressure sensors: an ion gauge, a convectron gauge, a dial gauge, and an absolute pressure transducer used to monitor the internal cryostat pressure.

\begin{figure}[t]
    \centering
    \includegraphics[width=\textwidth]{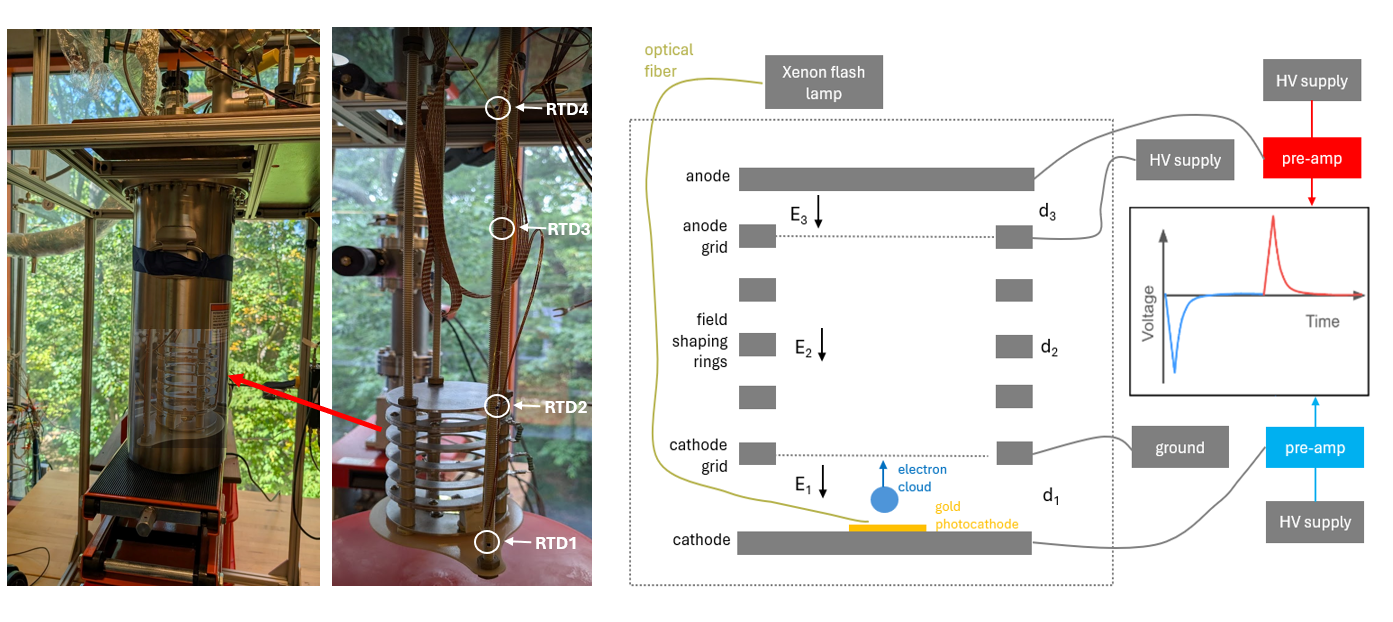}
    \caption{\label{fig:PrM}\textit{Left:} Cryostat with overlay showing the location of the UV-PrM. \textit{Middle:} UV-PrM and locations of the RTDs. \textit{Right:} Schematic diagram of the UV-PrM showing the optical and electrical connections.}
\end{figure}

As shown in Fig.~\ref{fig:PrM}, the UV-PrM is suspended from the top plate of the cryostat and fully immersed in LAr during operation. Four RTD sensors positioned at different heights in the chamber provide temperature and liquid level monitoring during operation. 
A baffle between the field cage and the top plate reduces thermal loading.

\begin{figure}[t]
    \centering
    \includegraphics[width=0.5\textwidth]{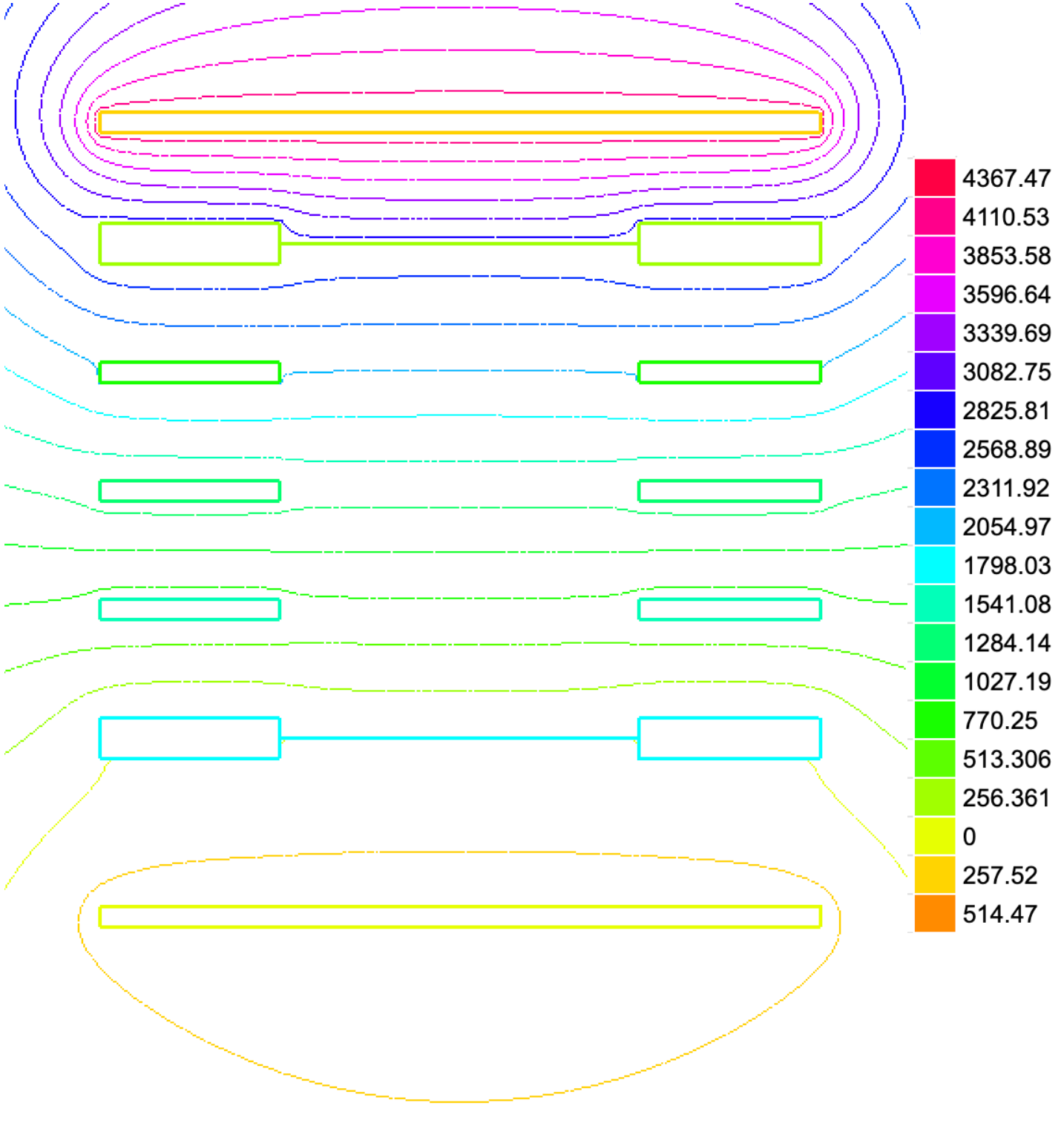}
    \caption{A 2D cross-section equipotential field map of the purity monitor made using the FreeFEM finite element software. The lines are color-coded based on the electric potential in volts.}
    \label{fig:FEM}
\end{figure}

The UV-PrM includes three regions of uniform electric field: between the cathode and cathode grid (cathode region), the cathode grid and anode grid (drift region), and the anode grid and anode (anode region), with respective lengths of 1.9\,cm, 5.8\,cm, and 1.2\,cm. 
Both grids are electroformed copper meshes (MC7 25LPI copper mesh by Precision Eforming) held in tension between two SS rings. Electric field uniformity in the drift region is maintained by three field-shaping rings connected by 50\,M\textohm{} resistors. The electrode geometry was simulated in the \textit{FreeFEM} finite element analysis software to verify the uniformity of the drift field within the active volume (see Fig.~\ref{fig:FEM}).

Electrons are produced at a photocathode via the photoelectric effect: ultraviolet light from a pulsed 60\,W xenon flash lamp (Hamamatsu L7685) is transmitted via optical fiber to a piece of gold-coated silicon attached (mechanically and electrically) to the cathode (see Fig.~\ref{fig:PrM}). The liberated electrons drift upward through the cathode grid toward the anode. The drifting electrons in the cathode region induce a current on the cathode. Likewise they induce a current on the anode electrode when drifting in the anode region. The grids electrically isolate the cathode and anode from induced signals when the electrons traverse the drift region. Appropriately large field ratios $E_3/E_2$ and $E_2/E_1$  are chosen to ensure that the grids are transparent to the drifting electrons \cite{fogarty_prm_2023}.   

The induced charge at the anode ($Q_A$) and cathode ($Q_C$) are measured using charge-sensitive preamplifiers (Cremat CR-110-R2). Any deficit in $Q_A$ relative to $Q_C$ is attributed to electron capture by impurities in the LAr, and the ratio $Q_A/Q_C$ provides a direct measure of the electron lifetime and impurity concentration in the LAr.

\begin{figure}[t]
    \centering
    \includegraphics[width=\textwidth]{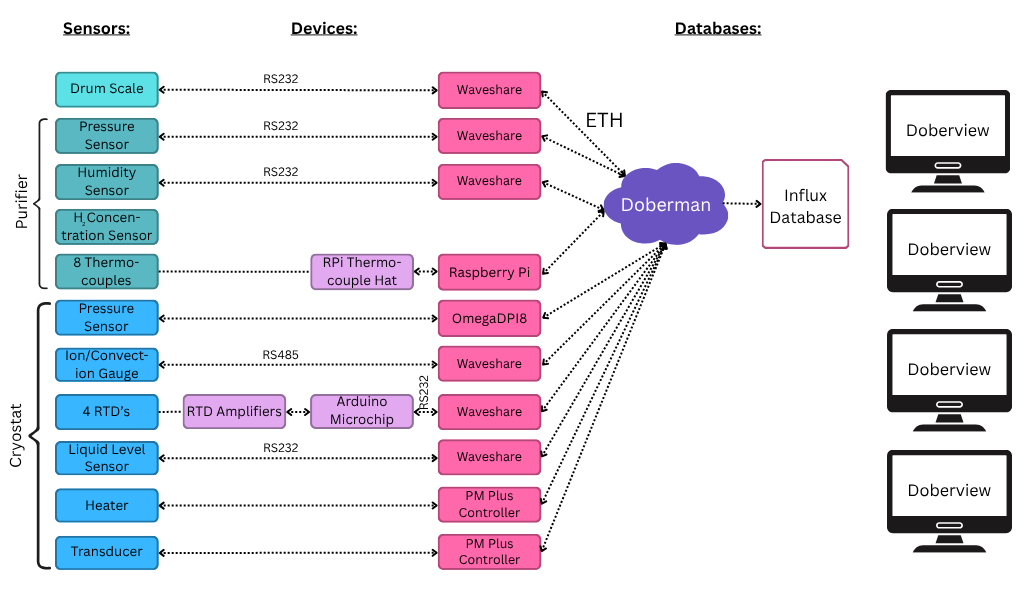}
    \caption{Diagram of slow control connectivity showing the interface devices between sensors and Doberman. Doberview, running on remote machines, is used to display slow control data.}
    \label{fig:SCDA_flow_chart}
\end{figure}

\subsection{Slow Control and Data Acquisition System}
\label{subsec:slow_control}
A slow control system monitors and records environmental and system operating parameters. The sensors include a drum scale to monitor the weight of LAr inside the supply dewar, a humidity sensor, pressure sensor, hydrogen concentration sensor, and thermocouples for the purifier, and pressure sensors and four RTDs for the cryostat. The sensors are connected to serial-to-ethernet adapters (Waveshare RS232/485 to ethernet) providing TCP/IP accessibility via static IP addresses. The Detector OBservation and Error Reporting Multiadaptive ApplicatioN (Doberman) software~\cite{dobermangit} handles the sensor communication and data logging to an Influx database. Data are viewable via the web using  Doberview~\cite{doberview} (see Fig.~\ref{fig:SCDA_flow_chart}). For the LAr measurements reported in this paper some sensors (purifier thermocouples and the H$_2$ monitor) were logged independently of Doberman.

\section{Experimental procedures}
\label{sec:procedure}

\subsection{Purifier regeneration}
\label{subsec:regeneration}

\begin{figure}[t]
    \centering
    \includegraphics[width=\textwidth]{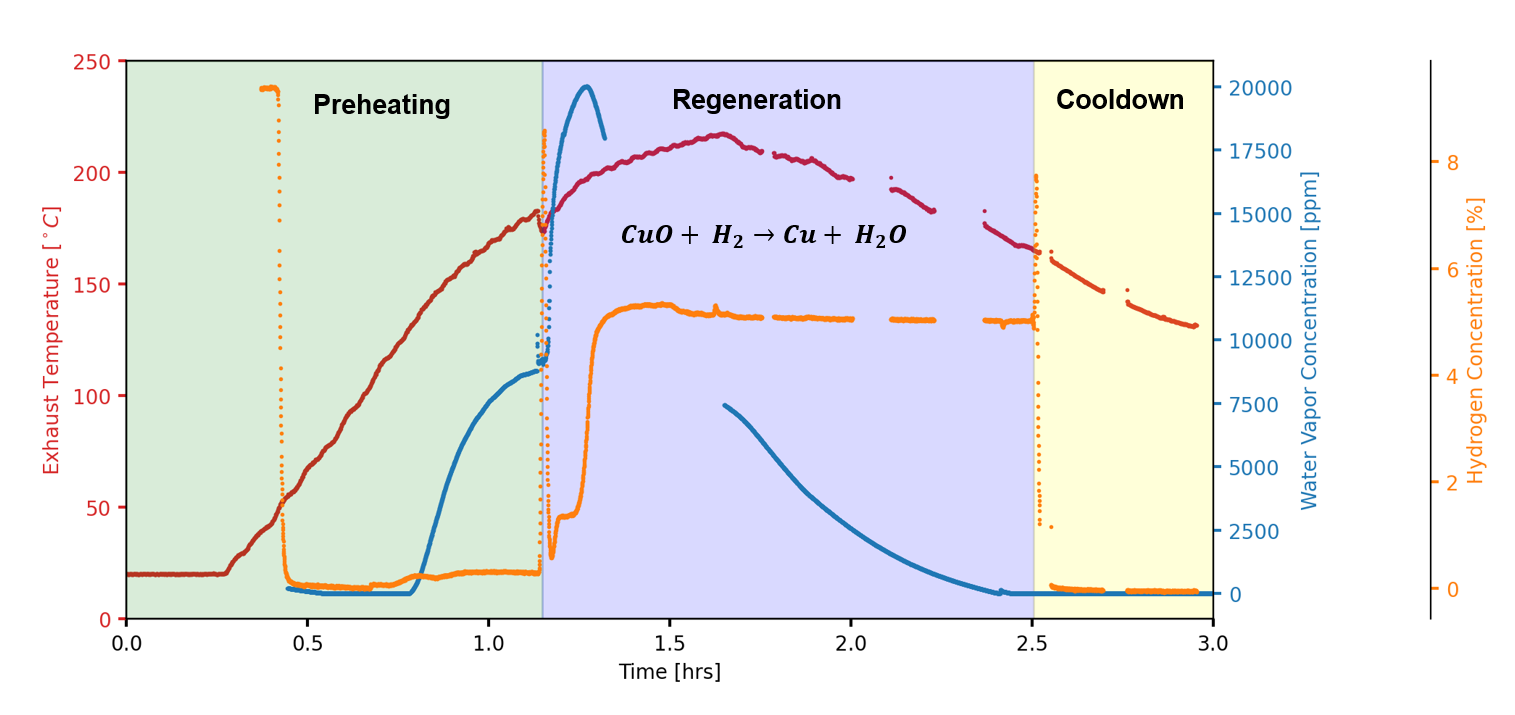}
    \caption{Temperature at the purifier outlet, along with water vapor and hydrogen concentration in the exhaust gas during filter regeneration. The cooldown stage continued for two hours beyond what is shown here. The gap in the water vapor data was caused by unexpected downtime of the Doberman slow-control system (at the time of this run, the temperature and water vapor data were logged separately from Doberman).}
    \label{fig:regen}
\end{figure}

The MS and AC filter media initially have significant adsorbed water and oxygen and therefore require initial regeneration prior to first use. The regeneration procedure (i) desorbs water from the MS and (ii) chemically reduces oxidized copper (CuO) to metallic copper (Cu) in the catalyst via the reaction:
\begin{equation}
    \mathrm{CuO} + \mathrm{H_2} \rightarrow \mathrm{Cu} + \mathrm{H_2O}.
\end{equation}
 The purifier was designed to filter 1000\,L of LAr before needing to be regenerated, though we typically perform regeneration one day prior to each LAr run to ensure optimal purifier performance. 
 
The purifier is regenerated through a three-stage procedure: preheating, hydrogen-assisted regeneration, and finally, cooldown and sealing. Figure~\ref{fig:regen} shows sensor data from a filter regeneration during all three stages of the process.

\begin{enumerate}
    \item \textbf{Preheating.} Ultra-high-purity (UHP) gas Ar is flowed through the purifier as the purifier is heated until the exhaust temperature reaches 180\,\textdegree{}C. During this stage water desorbs from the MS and is flushed out by the Ar flow.

    \item \textbf{Hydrogen-assisted regeneration.} 
    Once the target temperature is reached, the inlet gas is switched from UHP Ar to a Ar/H$_2$ mixture (95\%/5\%). The onset of hydrogen flow further increases the exhaust temperature due to the exothermic reduction of CuO. Likewise the exhaust gas humidity rises sharply since water is a byproduct of the reduction reaction. The Ar/H$_2$ flow rate is then reduced to ensure that the temperature of the filter media remains below 220\,\textdegree{}C to avoid damaging the AC. The MS and AC regeneration is considered complete once (i) the H$_2$ concentration in the exhaust gas returns to the input level (5\%), indicating that hydrogen is no longer being consumed in the reduction reaction and (ii) the water vapor concentration at the outlet decreases to zero indicating that the water has been desorbed from the MS.
    In our system, fully saturated filter media (as received from the vendor) requires about 12 hours for regeneration, while a substantially shorter cycle (15--30\,min) can "top up" recently recharged filter media.

    \item \textbf{Cooldown and sealing.} Once the regeneration stage is complete, the inlet gas is switched back to UHP Ar and the heaters are 
    turned off. UHP Ar is flowed at a high rate to cool the column, at which point the system is overpressurized by about 5\,psig with UHP Ar and sealed in preparation for future LAr purification.
\end{enumerate}

\subsection{LAr filling}
Before each LAr run, we do several pump-and-purge cycles to reduce residual impurities in the cryostat and liquid transfer lines. The system is first evacuated below 10$^{-4}$\,torr and then backfilled with UHP Ar to 5--7\,psig at room temperature. This sequence is repeated three to four times to dilute and remove contaminants released from internal surfaces. After the final purge, the chamber is pumped down once more and held under vacuum in preparation for the LAr fill.

Because the section of the liquid transfer line between the LAr storage dewar and purification column cannot be evacuated during the pump-and-purge cycles, it may contain residual contaminants. To minimize the transport of these contaminants into the chamber, the transfer line is purged through vents (near valves V1 and V13 in Fig.~\ref{fig:setup}). Once a steady liquid stream is observed at V13, the flow is redirected into the chamber.

During filling, the internal pressure of the cryostat is regulated by the pressure relief valve on the top plate. The liquid level inside the cryostat is monitored using the RTDs. Filling is stopped once all four RTDs show stable LAr temperatures, after which the pressure relief valve is sealed to prevent potential backflow contamination. 
The back-pressure regulator (with check valve) maintains a stable head pressure in the cryostat as the LAr slowly evaporates.
 
\section{System performance}
\label{sec:sys_perf}
The LAr purity is measured with a UV-PrM. 
A digital oscilloscope controlled by a custom Python script records waveforms from the cathode and anode charge-sensitive preamplifiers (CSP); the electron lifetime and LAr purity are determined as described below.

\subsection{Waveform acquisition and modeling}
\begin{figure}[t]
  \centering
  \includegraphics[width=\textwidth]{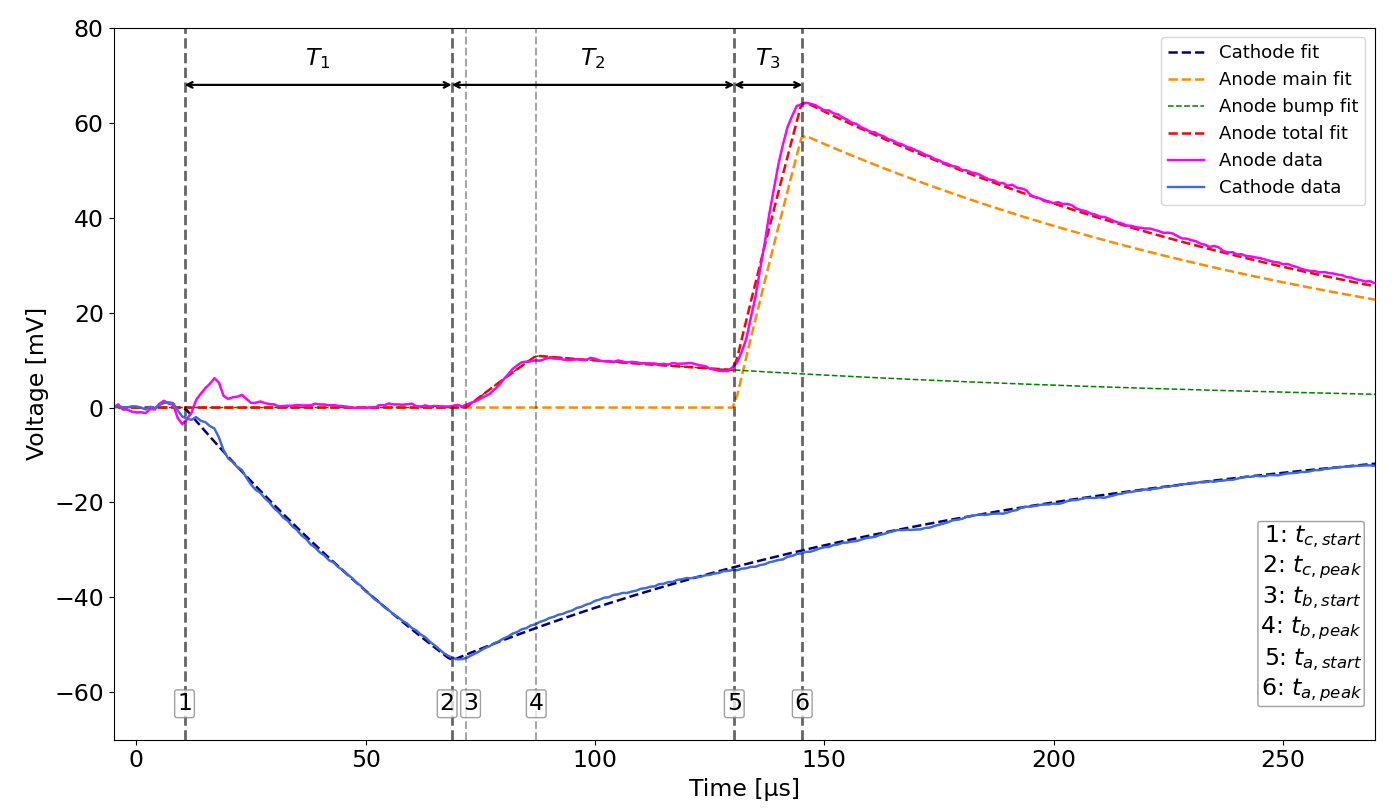}
  \caption{Example cathode and anode waveforms along with fitted model components. The drift times $T_1$, $T_2$, and $T_3$ are indicated on the plot. The relevant times for the two-component model used for the anode signal, as well as the times for the cathode signal are also indicated.}
  \label{fig:timing_diagram}
\end{figure}

During PrM operation, the xenon flash lamp repetition rate is typically 10\,Hz. To reduce noise in the CSP waveforms, we average hundreds of individual triggers on the oscilloscope and then record the averaged waveforms for offline analysis.
A typical pair of averaged cathode and anode waveforms is shown in Fig.~\ref{fig:timing_diagram}. The start of the rising edge of each pulse corresponds to the moment when the electrons enter the respective cathode or anode region (labeled "1" for the cathode waveform and "5" for the anode). The integral of the waveform is proportional to the total charge collected:
\begin{equation}
Q_C \propto \int V_C(t)\, dt,
\qquad
Q_A \propto \int V_A(t)\, dt.
\end{equation} 
We observe an unexpected feature (a "bump") in the anode waveform prior to the main anode peak. This feature is consistently present and has the characteristic CSP decay timescale, but it occurs too early to be generated by electrons in the anode region. We attribute it to parasitic capacitive coupling in the PrM. Our analysis accounts for the impact of this signal on the measurement of $Q_A$.

We define three time intervals (indicated visually in Fig.~\ref{fig:timing_diagram}):
\begin{itemize}
    \item $T_1 = t_\text{c,peak} - t_\text{c,start}$: electron drift time from the cathode to the cathode grid;
    \item $T_2 = t_\text{a,start} - t_\text{c,peak}$: electron drift time from the cathode grid to the anode grid;
    \item $T_3 = t_\text{a,peak} - t_\text{a,start}$: electron drift time from the anode grid to the anode.
\end{itemize}
The total electron drift time is then~\cite{Manenti:2020gzi}:
\begin{equation}
    t_{\mathrm{drift}} = T_2 + \frac{T_1 + T_3}{2},
    \label{eq:t_drift}
\end{equation}
which accounts for the average time electrons spend in the cathode ($T_1$) and anode ($T_3$) regions and the full drift time, $T_2$, in the drift region between the grids.

We model the cathode and anode waveforms as the convolution of the CSP impulse response function (an exponential decay with time constant $\tau_f$) with a constant current of duration $T$~\cite{Manenti:2020gzi}:
\begin{equation}
V_{\mathrm{out}}(t) =
\begin{cases}
h_0\,Q\,\dfrac{\tau_f}{T}\left(1 - e^{-t/\tau_f}\right),
& t < T, \\[8pt]
h_0\,Q\,\dfrac{\tau_f}{T}
\left(e^{T/\tau_f} - 1\right)
e^{-t/\tau_f},
& t \ge T.
\end{cases}
\label{eq:model}
\end{equation}
The decay time constant is the product of the resistance and capacitance in the feedback loop of the CSP: $\tau_f = R_f C_f$. Independent measurements of $\tau_f$ for the two CSPs provide \textit{a priori} constraints on those parameters (133\,$\mu$s and 135\,$\mu$s for anode and cathode, respectively). Our Cremat CR-110 CSPs include a secondary voltage amplification stage with a gain of two, so $h_0=\pm 2/C_f$ (with the minus sign for the cathode signal). We use a non-linear least-squares minimization to fit Eqn.~\ref{eq:model} to our data, with free parameters $Q$ (the total charge collected: $Q_C$ for cathode and $Q_A$ for anode), the "turn-on time" $t_{\mathrm{start}}$ ($t_{\mathrm{c,start}}$ for the cathode and $t_{\mathrm{a,start}}$ for the anode), and the duration of the current pulse $T$ ($T_1$ for the cathode and $T_3$ for the anode).
Our model of the anode data includes two components: one for the bump, and the second for the anode signal of interest:
\begin{equation}
V_A(t) = V_A^{\mathrm{anode}}(t - t_\text{a,start}) + V_A^{\mathrm{bump}}(t-t_\text{b,start}).
\end{equation}
In sum, our models for the cathode and anode waveforms are:
\begin{align}
V_C(t) &= V_C\!\left(
t, [\, t_{\mathrm{c,start}},\, t_{\mathrm{c,peak}},\, Q_{\mathrm{C}}] \right), \\
V_A(t) &= V_A\!\left(
t, [\, t_{\mathrm{a,start}},\, t_{\mathrm{a,peak}},\, Q_{\mathrm{A,anode}},\,
t_{\mathrm{b,start}},\, t_{\mathrm{b,peak}},\,
Q_{\mathrm{A,bump}}]\right),
\end{align}
where the terms in the square brackets are the fit parameters.

\subsection{Measured drift time, electron lifetime and impurity concentration}
We present three analyses using data from our PrM. The first is a consistency check comparing the measured and predicted electron drift times in the PrM. The second is a quantitative measurement of the LAr purity level over a 20-hour period of time. The third is, to our knowledge, a novel way to estimate the LAr purity by scanning the electric field strengths in the PrM. The main result is that with our single-pass purifier, we achieve sub-ppb oxygen-equivalent impurity concentration in the LAr, stable over the course of at least a day.

\begin{figure}[t]
  \centering
  \includegraphics[width=0.8\textwidth]{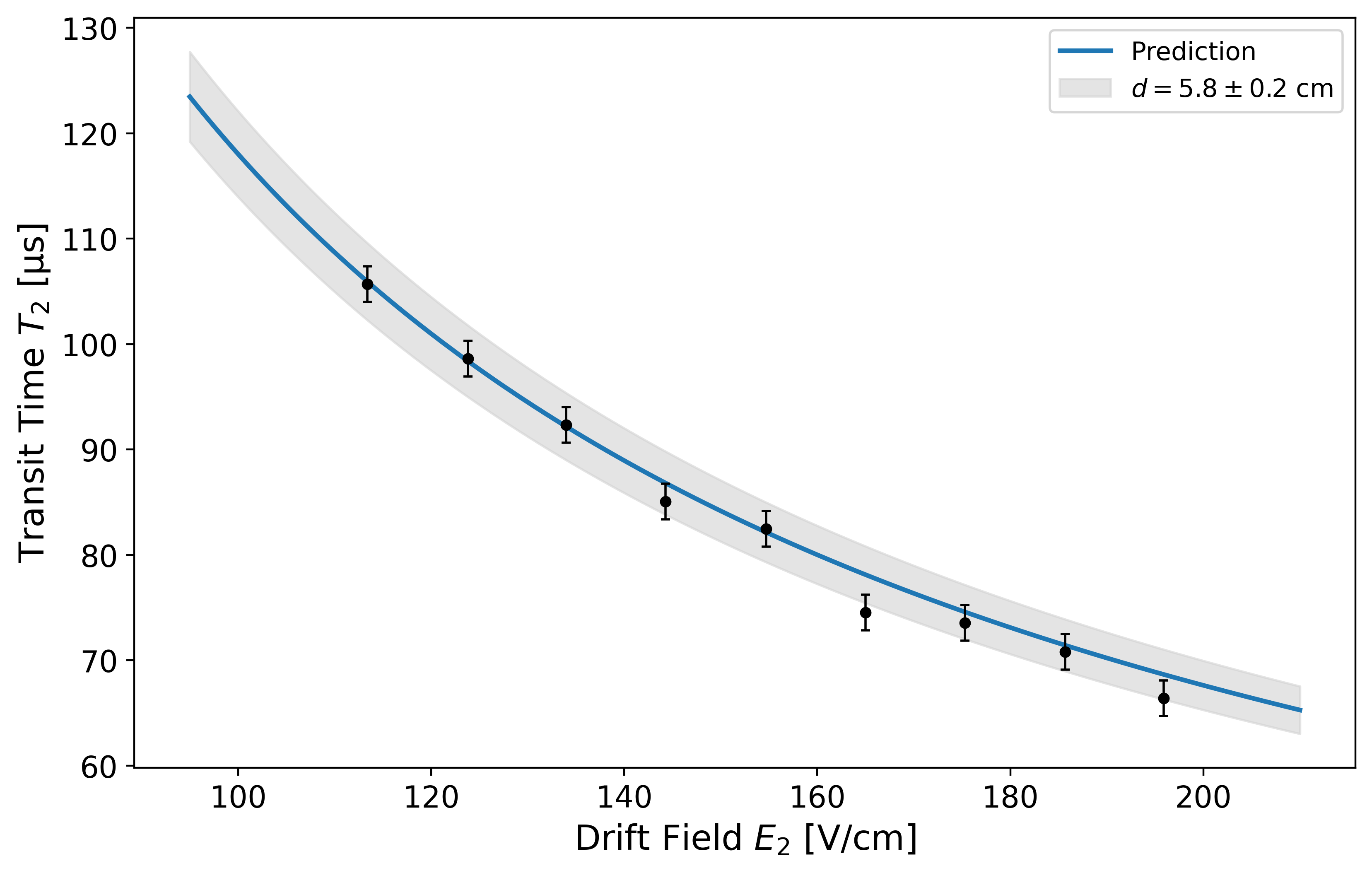}
  \caption{Measured and predicted drift region transit times $T_2$ as a function of the drift field $E_2$. The shaded band accounts for the systematic error due to a 2\,mm uncertainty in the drift distance.}
  \label{fig:vdrift_vs_E2}
\end{figure}

First we compare the measured and predicted transit times $T_2$ of the electrons in the drift region as a function of the drift electric field $E_2$. 
The predicted transit time is given by the ratio of the drift distance $d_2$ and the drift speed $v_{\mathrm{drift}}=\mu E_2$, where the mobility $\mu$ of electrons in LAr is itself dependent on both $E_2$ and the LAr temperature $T_{LAr}$~\cite{lar_properties}:
\begin{equation}
T_2 = \frac{d_2}{v_{\mathrm{drift}}} = \frac{d_2}{\mu(E_2,T_{LAr})E_2}.
\end{equation}
As seen in Fig.~\ref{fig:vdrift_vs_E2} we find good agreement between measured and predicted electron transit times. This reinforces the interpretation that the cathode and anode signals $V_{\mathrm{C}}(t)$ and $V_\mathrm{A}^{\mathrm{anode}}(t)$ are generated by electrons from the photocathode drifting through the cathode and anode regions, respectively. It also supports our claim that $V_\mathrm{A}^{\mathrm{bump}}(t)$ is not generated by electrons in the anode region. 

Next, we measure the electron lifetime $\tau$ directly from the measured charge survival fraction $Q_\mathrm{A,anode}/Q_C$ and electron drift time using~{\cite{Manenti:2020gzi}:
\begin{equation}\label{eq:tau}
\frac{Q_\mathrm{A,anode}}{Q_C} = e^{-t_\mathrm{drift}/\tau}.
\end{equation}
Notice that the numerator of the charge survival fraction does not include any contribution from the bump. We repeated this measurement over the course of one day (see Fig.~\ref{fig:tau_vs_time}), finding that the lifetime remained stable. The weighted average lifetime of these measurements is
\begin{equation}
\tau = (1.24\pm0.02)\,\mathrm{ms},
\end{equation}
where the reported uncertainty is the standard error on the mean.

\begin{figure}[t]
  \centering
  \includegraphics[width=1\textwidth]{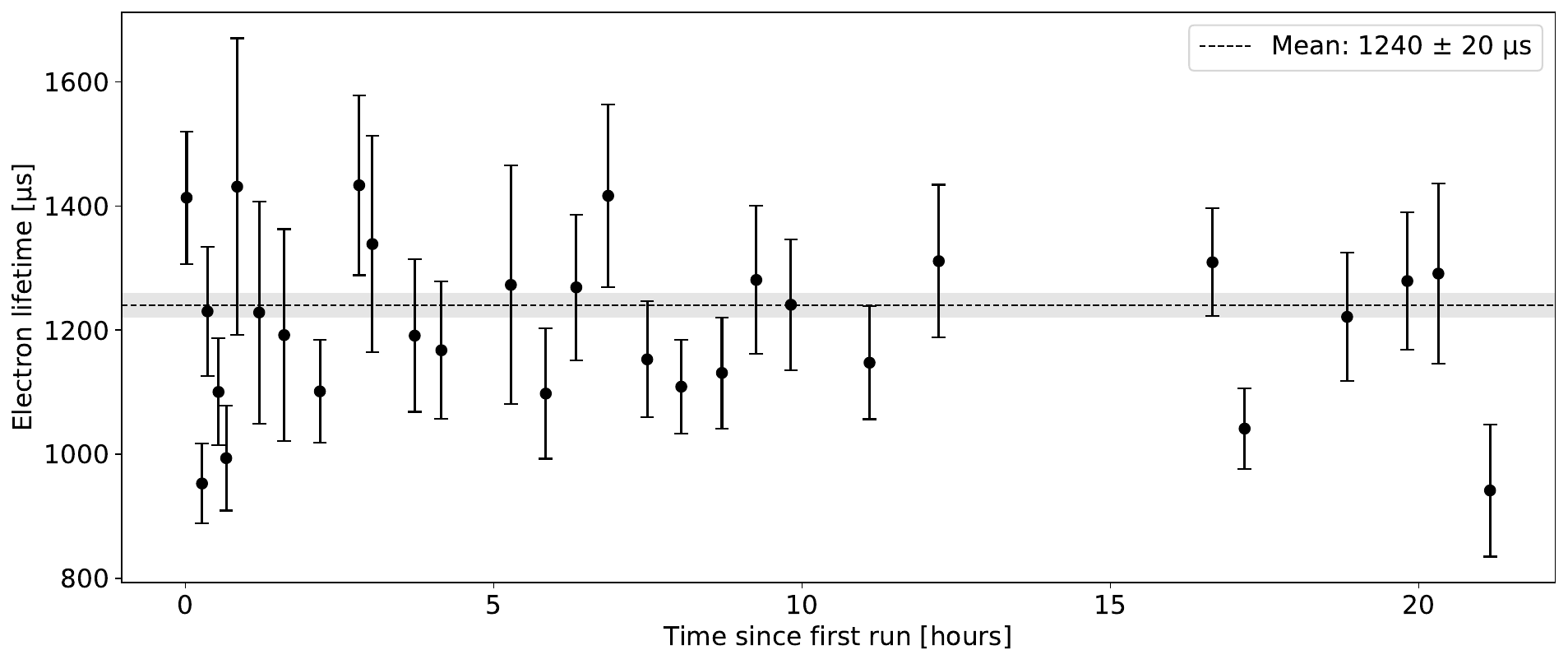}
  \caption{Electron lifetime over a 21-hour period with drift field $E_2 = 160$\,V/cm. The horizontal dashed line gives the weighted average lifetime over the full time span. The gray band indicates the standard error on the mean.}
  \label{fig:tau_vs_time}
\end{figure}

We then compute the LAr impurity concentration from the measured lifetime. From our measurements of $Q_A$ and $Q_C$ alone, we cannot disentangle the contributions of underlying impurity species (e.g. oxygen vs. water vs. nitrogen), so we follow standard practice and report the oxygen-equivalent impurity concentration $n_\mathrm{O_2}$ using the approach of Ref.~\cite{Li:2022pfu}:
\begin{equation}\label{eq:impurity}
n_\mathrm{O_2} = \frac{1}{k_\mathrm{O_2}\, \tau}, 
\end{equation}
where $k_\mathrm{O_2}$ is the impurity attachment coefficient of oxygen. The attachment coefficient depends on electric field. At our operating field of 160\,V/cm, we use 
%$k_\mathrm{O_2} = (4.11\pm 0.92)\times 10^{12}$\,s$^{-1}$, 
$k_\mathrm{O_2} = (3.84\pm 0.86)\times 10^{12}$\,s$^{-1}$, 
which comes from the rational polynomial parameterization (for impurity species $A$) from Ref.~\cite{Li:2022pfu}:
\begin{equation}\label{eq:ka}
k_A = 10^p\cdot \frac{\frac{a_1}{b_1} + a_1 E + a_2 E^2 + a_3 E^3 + a_4 E^4}{1 + b_1E + b_2 E^2 + b_3 E^3 + b_4 E^4},
\end{equation}
with $E$ the electric field in units of kV/cm (and $k_A$ in units of s$^{-1}$). The coefficients $p$, $a_i$ and $b_i$ for oxygen are also provided in Ref.~\cite{Li:2022pfu}.
% Note: the BNL web calculator uses old values: \cite{liPrivate}.
% ($a_1=39.4$, $a_2=1.20062$, $a_3=a_4=0$, $b_1=0.925794$, $b_2=1.63816$, $b_3=b_4=0$)
%In Li et al, the values were ($a_1=76.2749$, $a_2=4.24596$, $a_3=a_4=0$, $b_1=1.88083$, $b_2=2.62643$, $b_3=0.0632332$ and $b_4=-0.000211009$). 
The associated uncertainty on $k_\mathrm{O_2}$ is obtained from Fig.~1 of the same reference.
The resulting oxygen-equivalent impurity concentration is
\begin{equation}
n_\mathrm{O_2}=(0.21 \pm 0.05)\,\mathrm{ppb},
\end{equation}
where the error is predominantly due to the uncertainty in $k_\mathrm{O_2}$. 

Finally, we measure the charge survival fraction as a function of drift field $E_2$, which provides a semi-quantitative estimate of the electronegative impurity concentration in the LAr.
As we varied the drift field $E_2$ in this study, we also adjusted the cathode and anode electric fields $E_1$ and $E_3$ to maintain field ratios $E_3/E_2$ and $E_2/E_1$ above 2 for stable mesh transparency~\cite{Manenti:2020gzi}. Figure~\ref{fig:ratio_vs_E2} shows the result along with the predicted $Q_A/Q_C$ for a range of oxygen-equivalent impurity levels. The comparison suggests an  oxygen-equivalent impurity level of $0.2$ to $0.3$\,ppb, consistent with the impurity concentration measurement at $E_2=160$\,V/cm presented earlier.

\begin{figure}[t]
  \centering
\includegraphics[width=0.85\textwidth]{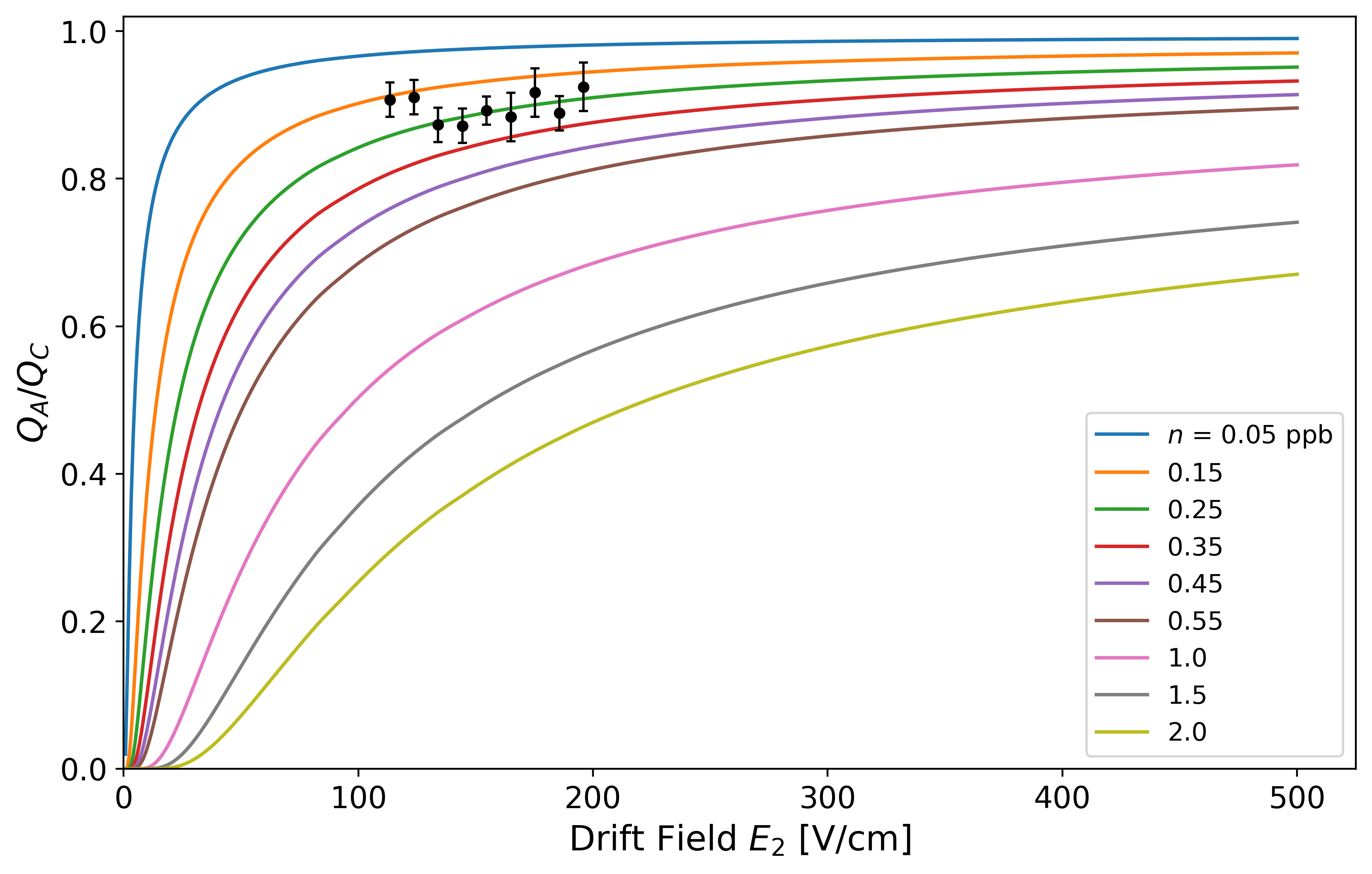}
  \caption{Charge ratio $Q_A/Q_C$ and theoretical predictions for O$_2$ concentrations as a function of the drift field $E_2$.}
  \label{fig:ratio_vs_E2}
\end{figure}

\section{Discussion and conclusion}
\label{sec:summary}
In summary, a compact 13-liter LAr test stand featuring a single-pass purifier and a UV-PrM has been successfully constructed and operated at Wellesley College. System performance was evaluated through drift-field scans and long-term purity measurements. The measured electron drift speeds are consistent with theoretical predictions, validating the UV-PrM configuration and timing analysis. The electron lifetime remained stable at 1.2\,ms ($E_2=160$\,V/cm) for more than 20~hours of continuous operation, demonstrating both the effectiveness of the purifier and the stable operation of the UV-PrM and cryogenic system. Data taking ultimately ended due to LAr boiloff, not degradation in purity. The measured electron lifetime corresponds to an O$_2$-equivalent impurity concentration of 0.2\,ppb, which yields an electron lifetime of 1.6\,ms at a drift field of 500\,V/cm. These results demonstrate that the system can produce stable sub-ppb impurity levels and millisecond-scale electron lifetimes.

Based on the experience gained with this 13-liter cryostat, we have built a larger version (250-liter), which employs the same UV-PrM design alongside a LArTPC to serve as platform for R\&D on Q-Pix and other LArTPC readout technologies. We have begun using the larger cryostat, fed by the same purification column used in this paper. Our first run achieved $n_\mathrm{O_2}=0.4$\,ppb. While we certainly expect to improve on that as we identify and eliminate impurity sources, this level of purity already exceeds our specification for the planned detector R\&D work as it corresponds to an electron drift distance of more than one meter at 500\,V/cm.

\acknowledgments
We thank Yichen Li (Brookhaven National Lab) and Yun-Tse Tsai (SLAC), as well as Michael Mooney and Sam Fogarty (Colorado State University) for their valuable support and discussions about purifier and purity monitor design and operation. We also thank Lindsay Donovan (Wellesley College) for her contribution to the measurements of the field cage. We acknowledge significant contributions by Larry Knowles (Wellesley College, deceased) to the mechanical design, machining and system assembly. This work was supported by the Gordon and Betty Moore Foundation (Grant No. GBMF11565 with DOI %\href{https://doi.org/10.37807/GBMF11565}{https://doi.org/10.37807/GBMF11565}
https://doi.org/10.37807/GBMF11565) and by the U.S. Department of Energy through the FAIR award DE-SC0024323.

% Bibliography
\bibliographystyle{JHEP}
\bibliography{biblio.bib}

\end{document}